\documentclass{aa}
\usepackage[varg]{txfonts}
\usepackage{graphicx}
\usepackage[colorlinks=true, allcolors=blue]{hyperref}
\usepackage{chemformula}

\newcommand{\be}{\begin{equation}}
\newcommand{\ee}{\end{equation}}

\usepackage{natbib}
\bibpunct{(}{)}{;}{a}{}{,}

\def\kHz{ \textrm{kHz} }
\def\MHz{ \textrm{MHz} }
\def\GHz{ \textrm{GHz} }

\newcommand{\hocs}{HOCS$^+$}
\newcommand{\hsco}{HSCO$^+$}

\begin{document}

\title{Advancing spectroscopic understanding of \ch{HOCS+}: Laboratory investigations and astronomical implications}
\author{Valerio Lattanzi\inst{1}, Miguel Sanz-Novo\inst{2}, V\'ictor M. Rivilla\inst{2}, Mitsunori Araki\inst{1}, Hayley A Bunn\inst{1}, Jes\'us Mart\'in-Pintado\inst{2}, Izaskun Jim\'enez-Serra\inst{2}  and Paola Caselli\inst{1}}

\institute{Center for Astrochemical Studies, Max-Planck-Institut f\"{u}r Extraterrestrische Physik, Gießenbachstraße~1, 85748 Garching, Germany \and Centro de Astrobiolog\'{i}a (CAB), INTA-CSIC, Carretera de Ajalvir km 4, Torrej\'{o}n de Ardoz, 28850 Madrid, Spain \\ \email{lattanzi@mpe.mpg.de}}

\date{\today}

\abstract{Sulphur-bearing species play crucial roles in interstellar chemistry, yet their precise characterisation remains challenging. Here, we present laboratory experiments aimed at extending the high-resolution spectroscopy of protonated carbonyl sulphide (\hocs), a recently detected molecular ion in space. Using a frequency-modulated free-space absorption spectrometer, we detected rotational transitions of \hocs\, in an extended negative glow discharge with a mixture of H$_2$ and OCS, extending the high-resolution rotational characterisation of the cation well into the millimetre wave region (200$-$370\,GHz). Comparisons with prior measurements and quantum chemical calculations revealed an overall agreement in the spectroscopic parameters. With the new spectroscopic dataset in hand, we re-investigated the observations of \ch{HOCS+} towards G+0.693$-$0.027, which were initially based solely on $K_a = 0$ lines contaminated by \ch{HNC^{34}S}. This re-investigation enabled the detection of weak $K_a\neq 0$ transitions, free from \ch{HNC^{34}S} contamination. Our high-resolution spectroscopic characterisation also provides valuable insights for future millimetre and submillimetre astronomical observations of these species in different interstellar environments. In particular, the new high-resolution catalogue will facilitate the search for this cation in cold dark clouds, where very narrow line widths are typically observed.}

\authorrunning{Lattanzi et al.}\titlerunning{New laboratory and astronomical insights on \hocs}
\maketitle

\section{Introduction}

The interstellar medium (ISM) stands as a vast cosmic laboratory, where a multitude of molecular species undergo complex chemical transformations under extreme conditions. Among the myriad of molecules detected in the ISM, sulphur-bearing species hold particular significance, comprising more than 10\% of the vast chemical inventory. These sulphur-containing compounds then play prominent roles in interstellar chemistry, influencing the formation of more complex molecules and contributing to the chemical diversity of the ISM. Despite the wealth of molecular data accumulated through astronomical observations, sulphur chemistry in the ISM remains a largely uncharted territory, especially in dense environments, where a significant depletion of gaseous sulphur onto interstellar dust grains is believed to occur \citep{Ruffle1999,laas2019modeling,Shingledecker2020}. Determining the sulphur content within both volatile and refractory compounds in the ISM poses a significant challenge in astrochemistry. The detection of primary sulphur reservoirs, icy H$_2$S and atomic gas, has proven difficult, relying heavily on the accuracy of models to estimate the abundances of species comprising less than 1\% of the total sulphur. Recent observations performed with JWST show that sulphur remains undepleted during the ionic, atomic, and molecular gas phases of the ISM, and in particular in the Orion Bar \citep{Fuente2024}. This is consistent with other findings that suggest that sulphur depletion is low in massive star-forming regions because of the interaction of the UV photons coming from newly formed stars with interstellar matter \citep{Fuente2023}. Similar low depletion has been observed towards the comet C67P/Churyumov-Gerasimenko \citep{Calmonte2016}. \\

Carbonyl sulphide, OCS, is one of the most abundant S-bearing species in the ISM, detected in a plethora of different environments, including pre-stellar cores such as L183 and L1544 \citep[e.g.][]{Lattanzi2020}, comets \citep{Bockelée-Morvan2016}, and interstellar ices \citep{Boogert2015}. An intriguing aspect of interstellar chemistry is the detection of many species that are unstable under terrestrial conditions, such as protonated species. The detection of protonated species in the ISM is of particular importance for several reasons. Firstly, these species provide valuable information about the chemical processes occurring in the interstellar environment, shedding light on the formation pathways and reaction mechanisms that govern molecular evolution. Secondly, protonated species often serve as intermediates in complex chemical reactions, acting as key players in the synthesis of more complex organic molecules. Lastly, in some cases, protonated species can serve as a proxy of their non-polar neutral counterparts, such as HOCO$^+$ and N$_2$H$^+$, of \ch{CO_2} and \ch{N_2}, respectively. Therefore, their detection can offer an insight into precursors of important pre-biotic molecules in space and their potential role in the emergence of life.\\

The most recent molecular ion detected in the interstellar gas is the oxygen-protonated carbonyl sulphide, \hocs, identified in the course of a molecular line survey of the molecular cloud G+0.693–0.027 (hereafter G+0.693) in the Galactic centre employing the Yebes 40m and IRAM 30m radio telescopes \citep{Sanz-Novo2024a}. While the sulphur protonated ion, \hsco\,, is more energetically favourable, by about 2500\,K \citep{Wheeler2006}, this isomer is yet to be identified in the ISM, leading to an estimated abundance of less than a factor of two compared to the high-energy \hocs\, isomer. Unfortunately, the $K_a = 0$ transitions of \ch{HOCS+} appeared to be blended with transitions from \ch{HNC^{34}S}, complicating the robust characterisation of the spectroscopic parameters of \ch{HOCS+}. Therefore, new laboratory experiments are needed to obtain information about the $K_a \neq 0$ transitions, which are not expected to be blended with those of \ch{HNC^{34}S}, and hence the spectroscopic information of \ch{HOCS+} can be properly determined.\\

The initial experimental investigation of the protonated carbonyl sulphide system was conducted by \citet{Nakanga1987}, during which the fundamental $\nu_1$ vibrational band (O-H stretch) of \hocs\, was detected via absorption in a hollow cathode discharge using a difference frequency laser system. Although these measurements were executed at a low resolution (with an accuracy of approximately 30 MHz), this study facilitated the inaugural experimental characterisation of the spectroscopic parameters of this cation. Concurrently, efforts to detect the lower energy isomer, \hsco, in the expected S-H stretch frequency region were unsuccessful. Approximately a decade later, \citet{Ohshima1996} advanced the understanding of the \hocs\, rotational spectrum in the centimetre-wave range, by detecting the lower three $K_a =0$ $a$-type transitions in the 11$-$35\,GHz range, through high-resolution measurements with a Fourier-transform microwave spectrometer coupled with a pulsed-discharge nozzle. The identification of the lower-energy isomer \hsco\ was ultimately achieved through the work of \citet{McCarthy2007}. Employing experimental techniques similar to those used by \citet{Ohshima1996} resulted in the detection of the three lowest rotational transitions in the $K_a=0$ ladder of the normal isotopic species, alongside DSCO$^+$ and H$^{34}$SCO$^+$. Further extensive  laboratory spectroscopic characterisation of \hsco\ and DSCO$^+$ extending into the submillimetre wave region was subsequently conducted by \citet{Lattanzi2018}.\\

Considerable scholarly attention has been directed towards the investigation of the geometries, vibrational wave numbers, and relative energies of two stable isomers of protonated carbonyl sulphide. Initial discrepancies between ab initio predictions and experimental results in infrared spectroscopy prompted an in-depth examination of the potential energy surface of this protonated system. This examination was meticulously carried out by \citet{Wheeler2006} employing multiple quantum chemical calculation methods, notably including coupled cluster theory extended to triple excitations, used in conjunction with the correlation consistent hierarchy of basis sets, cc-pVXZ (X=D,T,Q,5,6). Subsequently, comprehensive ab initio spectroscopic predictions, encompassing fundamental vibrational frequencies and spectroscopic constants of \hocs, \hsco, and their isotopically substituted variants, were undertaken by \citet{Fortenberry2012}. These studies utilised vibrational perturbation theory at the second order and the vibrational configuration interaction method, integrated with an extremely accurate quartic force field. Furthermore, an exhaustive review of high-level quantum chemical calculations concerning several sulphur-containing species, including protonated carbonyl sulphide, was conducted by \citet{Alessandrini2018}. This review employed the coupled-cluster singles and doubles approach (CCSD) augmented by a perturbative treatment of triple excitations, denoted as CCSD(T).\\

In this manuscript, we present the results of laboratory experiments aimed at extending the high-resolution spectroscopy of \hocs\,, providing a reliable database for the astronomical characterisation of this species well into the submillimetre wave region, and including a full treatment of the asymmetric rotor behaviour. A search for $K_a \neq 0$ $a$-type transitions in G+0.693 is also presented and discussed in the following sections, along with a re-investigation of the $K_a=0$ transitions.

\section{Laboratory experiment}

The laboratory experiment utilised the Center for Astrochemical Studies Absorption Cell (CASAC), a frequency-modulated free-space absorption spectrometer, developed at the Max Planck Institute for Extraterrestrial Physics \citep{Bizzocchi2017}. This instrument has previously been used to characterise various reactive sulphur-bearing species \citep[e.g.][]{Prudenzano2018,Lattanzi2018,Inostroza-Pino2023,Araki2024}. The primary radiation source is a Keysight E8257D frequency synthesiser precisely synchronised with a 10\,MHz rubidium frequency standard from Stanford Research Systems, ensuring the highest accuracy in frequency and phase stabilisation. The synthesiser's radiation can then be amplified and multiplied using a Virginia Diodes solid-state active multiplier chain, allowing the coverage of the 75\,GHz to 1600\,GHz frequency range with exceptional frequency agility.  

Radiation passes through a 3-metre-long, 5-centimetre-diameter Pyrex tube, encountering two hollow stainless steel electrodes, each 10 centimetres in length, connected to a 5\,kW DC power supply. The discharge region, which spans 2 metres between the electrodes, can be efficiently cooled by liquid nitrogen. Frequency modulation is achieved by encoding the signal with a sine wave at a constant rate of 50\,kHz. Upon interaction with the molecular plasma, the signal is detected using a cryogenic-free InSb hot electron bolometer from QMC Instruments Ltd. To derive the absorption signal's second derivative profile, a lock-in amplifier (SR830, Stanford Research Systems) demodulates the detector output at twice the modulation frequency (2f detection). 

All experimental procedures were co-ordinated and recorded using a computer-controlled acquisition system. The experimental conditions involved the gases Neon, H$_2$, and OCS in a ratio of 30/1/1, respectively, and for a total pressure of 20\,mTorr measured downstream. The whole cell was cooled using liquid nitrogen to a temperature of $\sim$\,150\,K, measured around the middle of the length of the Pyrex tube, and an extended negative glow discharge was applied at 2\,kV/6\,mA with a $\sim$\,200\,G co-axial magnetic field. 

Overall, the absorption characteristics exhibited weak signals, which required extensive averaging to achieve an acceptable signal-to-noise ratio. Undoubtedly, temperature emerged as the most crucial factor in ensuring a satisfactory quality of signal readability. In an effort to amplify the weak signal, additional tests were conducted using a hollow cathode system recently developed in our laboratory. However, these attempts proved unsuccessful. Detailed findings regarding this new system will be presented separately. The general experimental conditions were quite akin to those used for the spectroscopy of the lower-energy isomer and reported in \cite{Lattanzi2018}, with the sole exception of the buffer gas; in this case, Neon demonstrated superior discharge stability and facilitated better dynamic control of the experiment, while in the case of \hsco\, measurements Argon was preferred.

\begin{figure}[htbp]
    \centering
    \resizebox{\hsize}{!}{\includegraphics{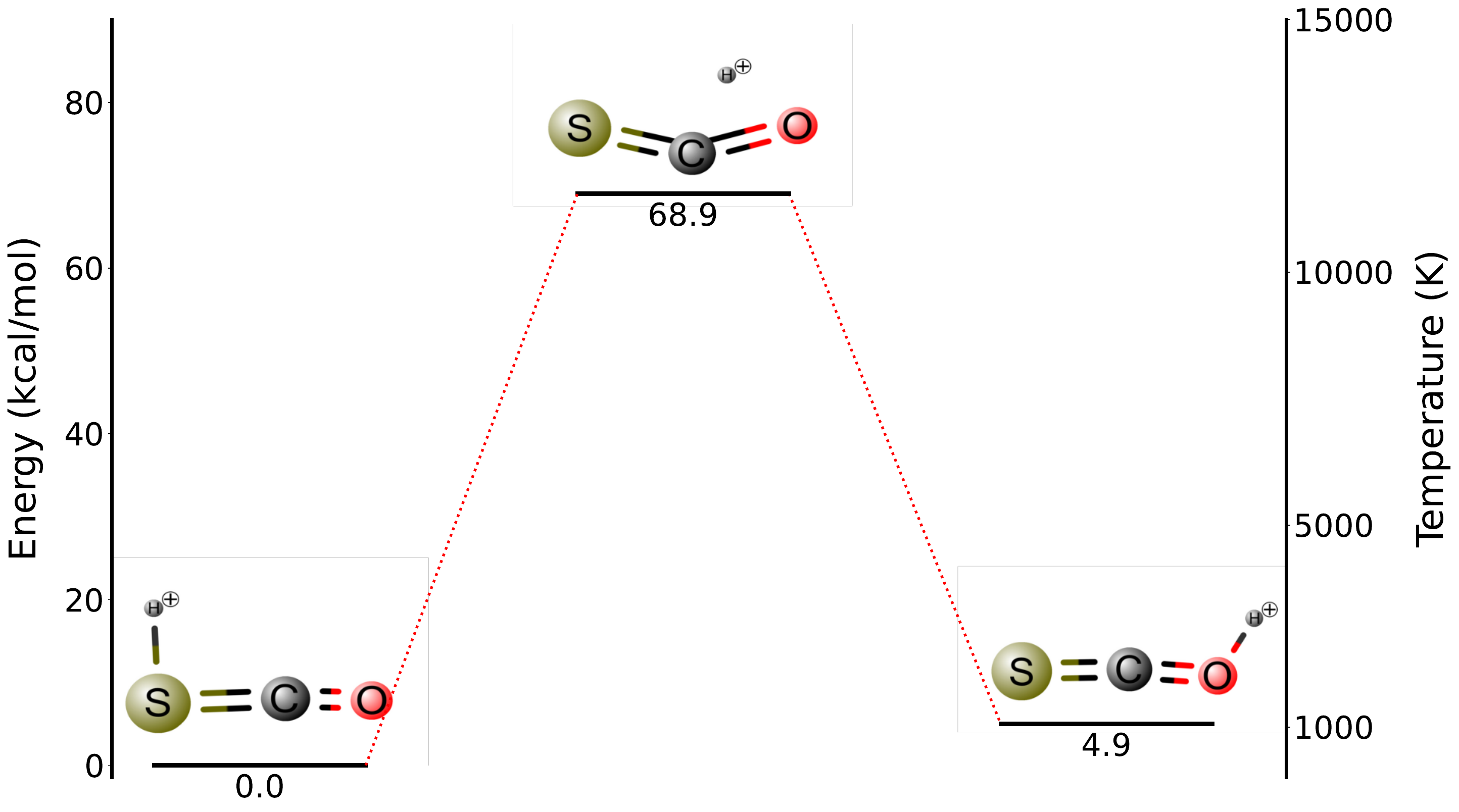}}
    \caption{Structures and relative energy levels of \hsco, \hocs, and the \ch{SC(H)O+} transition state. Values refer to kcal/mol and have been obtained at the CCSD(T) level of theory, with the cc-pVQZ basis set by \cite{Wheeler06}.} 
    \label{fig:pes}
\end{figure}

\section{Analysis}\label{sec:analysis}

The protonated carbonyl sulphide system comprises two stable isomers, depending on whether the protonation occurs to the sulphur or the oxygen side. The energy separation of the two species is about 4.9\,kcal/mol ($\sim$\,2450\,K), with the sulphur protonated isomer being the lower-energy form. The transition state, which constitutes the barrier to the isomerisation of \hsco\, to \hocs, lies $\sim$\,69\,kcal/mol higher than the lower-energy isomer (see Fig.\,\ref{fig:pes}). The oxygen-protonated carbonyl sulphide is a near-prolate asymmetric top ($\kappa=-0.99$) with a $C_s$ symmetry and a planar structure (inertial defect of $\Delta \approx 0.082$\,amu\,$\AA^2$). Proton affinities of OCS to form \hocs\, and \hsco\, were calculated to be 610\,kJ\,mol$^{-1}$ and 632\,kJ\,mol$^{-1}$, respectively, with the CCSD(T)/aug-cc-pVTZ method \citep{Tsuge2016}. \\

The laboratory search was motivated by the recent detection of \hocs in the G+0.693 molecular cloud by \citet{Sanz-Novo2024a} and by the previous successful experiment carried out in our laboratories on \hsco. Assuming similar experimental conditions for the formation of both species, the experimental settings were first optimised for the production of the previously detected \hsco. Once the conditions were defined for the S-protonated OCS, a search was carried out for the oxygen-protonated isomer around the strongest predicted $a$-type $K_a = 0$ transitions in the 1mm-wave band. This frequency region was chosen as a good balance between the predicted line intensity for our experimental conditions and the overall performance of the spectrometer. The initial search was guided by the effective $B$ rotational and quartic distortion constants derived by \citet{Ohshima1996}.

\begin{figure*}[htbp]
    \centering
    \includegraphics[scale=0.13]{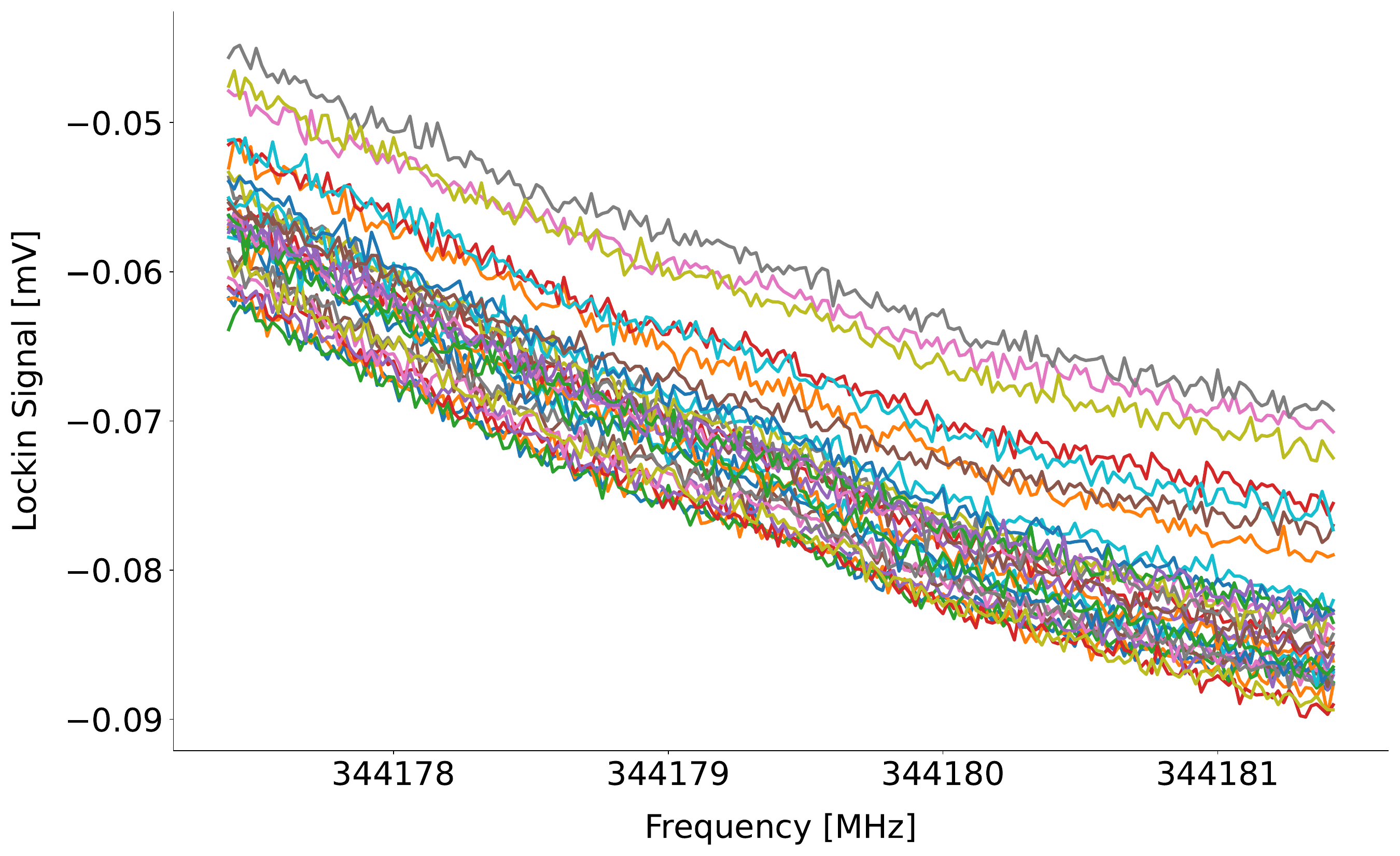}
    \includegraphics[scale=0.13]{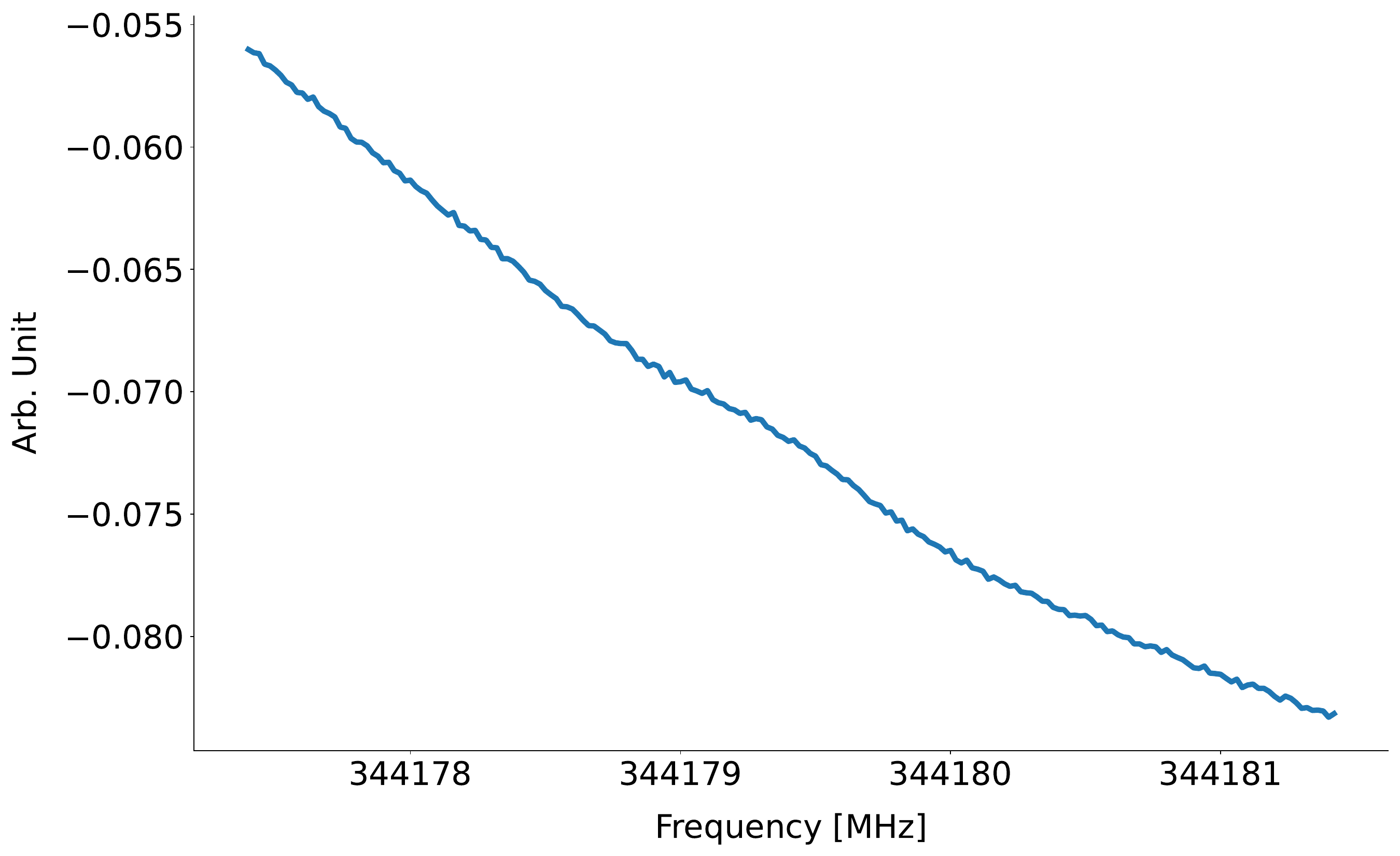}
    \includegraphics[scale=0.13]{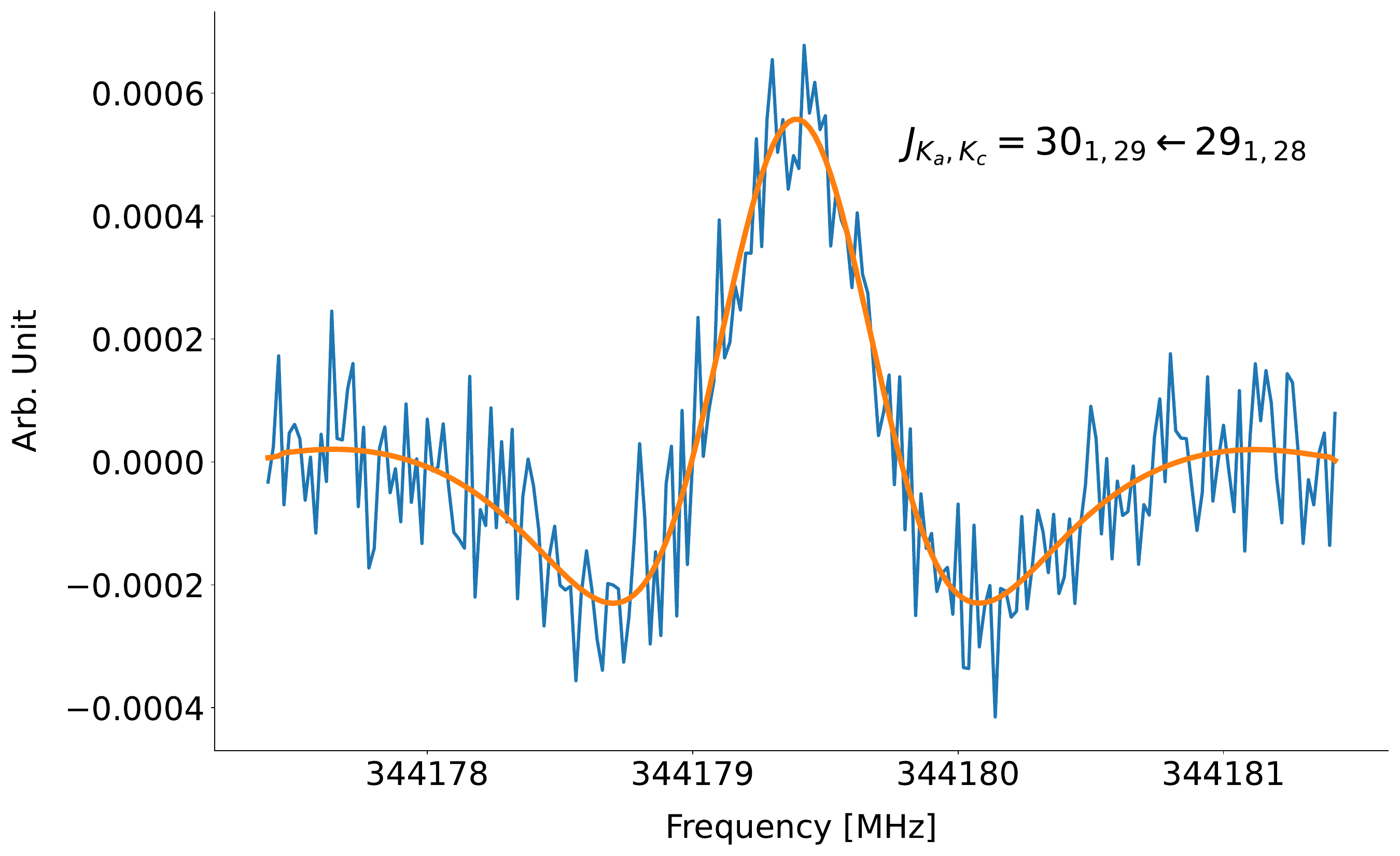}
    \caption{Experimental rotational spectra of \hocs. \textit{Left panel}: Single short scans (3-minute integration time, each with a 3 ms time constant) around 344\,GHz. \textit{Centre panel}: Average spectrum from the 30 individual scans. \textit{Right panel}: $2f$ absorption spectrum of the $J_{K_a,K_c} = 30_{1,29}-29_{1,28}$ rotational transition obtained by removing the baseline from the spectrum in the centre panel (\textit{blue}) and the best fit to a speed-dependent Voigt profile (\textit{orange}) (see text).}
    \label{fig:spectra}
\end{figure*}

Multiple short integrations on the same transition, as opposed to a single long intergration, were required due to the delicate and sensitive experimental setting. With this approach, a better handling on the stability of the experimental conditions was possible, along with the opportunity to monitor bad scans that were eventually discarded from the final averaging. The results of such a strategy can be appreciated in Fig.\,\ref{fig:spectra} where the rotational frequency of the transition, $J_{K_a,K_c} = 30_{1,29}-29_{1,28}$, around 344179\,MHz, has been obtained after averaging 30 different scans, each with an integration time of 3 minutes. Although the individual averaging periods were brief, one can observe significant variability in the experimental conditions, as is evidenced by notable fluctuations in the baseline (Fig.\,\ref{fig:spectra}, top). This variability vividly illustrates the interplay among various experimental parameters within a discharge system. Even slight fluctuations in cell temperature, spanning just a few degrees, exert a discernible impact on plasma behaviour, thereby influencing the efficacy of the DC discharge and the applied magnetic field. Concurrently, alterations in discharge and magnetic field conditions reciprocally affect the molecular plasma and its effective temperature. To recover the line central frequencies, each individual experimental averaged rotational transition was baseline-subtracted and line-profile-fitted to a modulated Voigt algorithm \citep{Dore2003}, implemented in the data analysis software QtFit, part of our in-house python-based libraries for laboratory spectroscopy, pyLabSpec.\footnote{https://laasworld.de/pylabspec.php} Both the complex component of the Fourier transform of the dipole correlation function (i.e. the dispersion term) and a low-order polynomial (typically second or third order) were also taken into account to model the line asymmetry and baseline produced by the background standing waves between non-perfectly transmitting windows of the absorption cell.
From this analysis, considering the line width and signal-to-noise ratio of each rotational transition, we assigned an estimated accuracy of 75--150\,kHz to our experimental dataset. With the line central frequency in hand, the whole dataset, including the three low-frequency transitions previously measured, was then analysed using the SPFIT/SPCAT suit of programs \citep{Pickett1991} and fitted to a Watson S-reduced Hamiltonian for asymmetric-top molecules. The full dataset now consists of 30 rotational transitions, from 28 independent absorption features, including the three centimetre-wave transitions previously detected by \citet{Ohshima1996}, and a final standard deviation of 46\,kHz (see Table\,\ref{table:1}). The transitions observed by \citet{Nakanga1987} were not included in the final dataset since they were not available and also measured with a reported accuracy of 0.001\,cm$^{-1}$ (i.e. $\sim$\,30\,MHz), which is at least 200 times worse than our estimated accuracy. Nevertheless, it is worth comparing the molecular parameters derived in the new dataset with those reported earlier, including the infrared measurements of the $\nu_1$ fundamental band by \citet{Nakanga1987} and the more recent ab initio quantum chemical calculations by \citet{Alessandrini2018} (see Table\,\ref{table:1}). An initial comparison of the spectroscopic parameters obtained in the different works shows a good agreement amongst the majority of parameters. In particular, the theoretical values for $B$ and $C$ are only a few hundred kilohertz away from the experimental values (0.02\% and 0.01\%). The quartic centrifugal distortion constants, $D_J$ and $D_{JK}$, both agree within 3\%. The largest difference is seen for the $d_1$ quartic parameter;
here, the difference between the parameter derived experimentally and that obtained from quantum chemical calculations is around a few tens of percent. This difference is even larger when the ab initio is compared to the previous experimental value obtained by \citet{Nakanga1987}. For consistency with our work and other ab initio and experimental studies, the parameters from \citet{Nakanga1987} shown in Table\,\ref{table:1} have been converted from the Watson A-reduction Hamiltonian used by the authors in their work to the Watson S-reduced one, following Table\,8.16 of the \citet{Gordy1984}. It is worth noticing that the real impact on this conversion is marginal since the magnitude of the $d_2$ parameter is negligible (less than 1\,Hz) especially when compared to all the other parameters, and the only visible effect is on the $d_1$ constants, which change sign when passing from $\delta_J$, since $d_1 = -\delta_J$. Due to the limitation of our measurements to the $a$-type spectrum only, the $A$ rotational constant is still largely uncertain ($\sim$\,0.5\% error), compared to the other rotational constant, but in good agreement with the theoretical value (0.02\% difference), and in one sigma from the previous experimental value. The other expected effect of the limitation to the $a$-type spectrum is on the large unresolved correlation of the $A$ and $D_K$ parameters, so the latter is left fixed to the measured values by \citet{Nakanga1987}. An alternative fitting analysis was performed, fixing the $D_K$ to its ab initio value derived in \citet{Alessandrini2018}, but no significant variations were noticed in the overall quality of the fit itself and the values, and corresponding errors, of the other individual parameters.

\begin{table*}[htbp]
\caption{Spectroscopic parameters of oxygen protonated carbonyl sulphide \hocs.}
\label{table:1} \centering
\begin{tabular}{llr@{.}lr@{.}lr@{.}lr@{.}l}
\hline\hline   
\noalign{\smallskip}
Parameter & unit & \multicolumn{2}{c}{This work} & \multicolumn{2}{c}{\citet{Nakanga1987}$^a$}  & \multicolumn{2}{c}{\citet{Ohshima1996}$^b$}  &\multicolumn{2}{c}{ab initio$^c$} \\
\hline
\noalign{\smallskip}
   $A$      & \GHz  &        779&1(41)         &      782&6957(38)      &  \multicolumn{2}{c}{}  &   779&260      \\
   $B$      & \MHz  &        5750&3771(30)     &     5750&551(49)       &     5726&66011(10)       &  5749&14       \\
   $C$      & \MHz  &        5702&9444(32)     &     5703&030(50)       &  \multicolumn{2}{c}{}  &  5702&15       \\
   $D_J$    & \kHz  &           1&04165(45)    &        1&104(20)       &        1&0640(80)       &     1&01       \\
   $D_{JK}$ & \kHz  &         261&19(13)       &      281&9(15)         &  \multicolumn{2}{c}{}  &   270&         \\
   $D_K$    & \MHz  &         993&68$^d$       &      993&68(77)        &  \multicolumn{2}{c}{}  &   649&01       \\
   $d_1$    & \kHz  &          $-$0&01037(99)    &      $-$0&0198(43)      &  \multicolumn{2}{c}{}  &    $-$0&00706    \\
   $d_2$    & \kHz  &          $-$0&000833$^e$   &  \multicolumn{2}{c}{}  &  \multicolumn{2}{c}{}  &    $-$0&000833   \\
\hline
\noalign{\smallskip}
\# lines       &      & \multicolumn{2}{c}{30} \\
$\sigma_{rms}$ & \kHz & \multicolumn{2}{c}{46} \\  
$\sigma_{w}^f$ &      & \multicolumn{2}{c}{0.63}\\
\hline
\end{tabular}
\tablefoot{Values in parentheses represent 1$\sigma$ uncertainties, expressed in units of the last quoted digit.\\
$^a$Parameters converted from Watson A-reduction Hamiltonian to S-reduction (see text). \\
$^b$Values reported in \citet{Ohshima1996} are actually $B_{eff} = (B + C)/2$ and $D_{eff} = D_J + (B-C)^2/\{32[A-(B +C)/2]\}$.\\
$^c$\citet{Alessandrini2018}.\\
$^d$ Fixed to the value derived in \citet{Nakanga1987}.\\
$^e$ Fixed to the ab initio value. \\
$^f$Dimensionless rms, defined as $\sigma_{w} = \sqrt{\frac{\sum_i\left(\delta_i/err_i\right)^2}{N}}$, where the $\delta$s are the residuals weighted by the experimental uncertainty (\emph{err}) and \emph{N} the total number of transitions analysed.}
\end{table*}

\section{Astronomical search} 
\label{sec:astro}

\subsection{Observations} 
\label{subsec:obs}

Our search for \ch{HOCS+} involved an unbiased and ultra-deep spectral survey performed towards the Galactic centre molecular cloud G+0.693. This source ranks among the most chemically diverse astronomical regions, marked by the first detections of numerous carbon-, oxygen-, nitrogen-, and sulphur-bearing compounds (see, e.g. \citealt{rivilla2019b,rivilla2020b,rivilla2021b,rivilla2021a,rivilla2022a,rivilla2022b,Rivilla23,rodriguez-almeida2021a,rodriguez-almeida2021b,jimenez-serra2022,zeng2021,zeng2023,SanzNovo23,Sanz-Novo2024a,Sanz-Novo2024b}).

We covered the $Q$-band (31.075$-$50.424 GHz) using the Yebes 40$\,$m (Guadalajara, Spain) radiotelescope. Furthermore, the IRAM 30\,m radiotelescope was employed to cover three additional frequency ranges with high sensitivity: 83.2$-$115.41\,GHz, 132.28$-$140.39\,GHz, and 142.00$-$173.81\,GHz. We used the position-switching mode, centred at $\alpha$ = $\,$17$^{\rm h}$47$^{\rm m}$22$^{\rm s}$, $\delta$ = $\,-$28$^{\circ}$21$^{\prime}$27$^{\prime\prime}$, with the off position shifted by $\Delta\alpha$~=~$-885$$^{\prime\prime}$ and $\Delta\delta$~=~$290$$^{\prime\prime}$. The half power beam width (HPBW) of the Yebes 40$\,$m telescope ranges from approximately 35$^{\prime\prime}$ to 55$^{\prime\prime}$ at frequencies of 50 GHz and 31 GHz, respectively \citep{tercero2021}. Meanwhile, the HPBW of the IRAM 30$\,$m radiotelescope varies from $\sim$14$^{\prime\prime}$ to $\sim$29$^{\prime\prime}$ across the covered frequency range. Further details of these observations (e.g. resolution, beam efficiencies, and noise levels of the molecular line survey) are provided in \citet{Rivilla23} and \citet{SanzNovo23}.

\begin{table*}
\centering
\tabcolsep 3pt
\caption{Spectroscopic information of the selected $K_{a}$ = 0 and 1 transitions of \ch{HOCS+} detected towards G+0.693$-$0.027 (shown in Fig. \ref{f:LTEspectrum}).}
\begin{tabular}{r@{.}lr@{\,--\,}lccccccccc}
\hline\hline
\multicolumn{2}{c}{Frequency} & \multicolumn{2}{c}{Transition} $^{(a)}$ & log \textit{I}& \textit{g}$\mathrm{_u}$ & $E$$\mathrm{_{up}}$ &  Blending  \\ 
\multicolumn{2}{c}{(GHz)} & \multicolumn{2}{c}{} &  (nm$^2$ MHz) &  &  (K)  & \\
\hline
\noalign{\smallskip}
34&3598431*  & 3$_{0,3}$ & 2$_{0,2}$  & --5.0117  & 7  & 3.3 &  \ch{HNC$^{34}$S} \\ 
45&8129972*  & 4$_{0,4}$ & 3$_{0,3}$  & --4.6397  & 9  & 5.5   & \ch{HNC$^{34}$S} \\
80&1716982*  & 7$_{0,7}$ & 6$_{0,6}$  & --3.9237  & 15  &  15.3 & \ch{HNC$^{34}$S} and \ch{\textit{t}-C2H3CHO} \\
91&6242541*  & 8$_{0,8}$ & 7$_{0,7}$  & --3.7557  & 17  & 19.7 & \ch{HNC$^{34}$S} \\
103&0765925*  & 9$_{0,9}$ & 8$_{0,8}$  & --3.6090  & 19 &  24.6 & \ch{HNC$^{34}$S} and \ch{N-CH3NHCHO}\\
114&5286862*  & 10$_{0,10}$ & 9$_{0,9}$  & --3.4793  & 21  &  30.0 & \ch{HNC$^{34}$S} \\
125&9805079*  & 11$_{0,11}$ & 10$_{0,10}$  & --3.3635  & 23 &  36.0  & \ch{HNC$^{34}$S} and \ch{H$^{13}$CONH2}  \\
137&4320305*  & 12$_{0,12}$ & 11$_{0,11}$  & --3.2593  & 25  & 42.6 & \ch{HNC$^{34}$S} \\
148&8832268*  & 13$_{0,13}$ & 12$_{0,12}$  & --3.1649  & 27  & 49.7 & \ch{HNC$^{34}$S} and \ch{HCOCH2OH} \\
160&3340697*  & 14$_{0,14}$ & 13$_{0,13}$  & --3.0791  & 29  & 57.3 & \ch{HNC$^{34}$S} and \ch{\textit{Z}-CH3CHNH} \\
\hline 
 34&4294313  &   3$_{1,2}$ & 2$_{1,1}$    & --5.1147  & 7   & 40.1 &  Unblended \\ 
 45&9057880  &   4$_{1,3}$ & 3$_{1,2}$    & --4.7196  & 9   & 42.3 &  CH$_3$$^{13}$CH$_2$CN and U-line \\ 
 80&3341357  &   7$_{1,6}$ & 6$_{1,5}$    & --3.9846  & 15  & 52.1 &  U-line \\ 
 91&4305056  &   8$_{1,8}$ & 7$_{1,7}$    & --3.8180  & 17  & 56.4 &  Unblended \\ 
 91&8099247  &   8$_{1,7}$ & 7$_{1,6}$    & --3.8145  & 17  & 56.5 &  \ch{CH3CONH2} \\ 
102&8586736  &   9$_{1,9}$ & 8$_{1,8}$    & --3.6698  & 19  & 61.3 &  \ch{H2NC(O)NH2}\\ 
103&2855074  &   9$_{1,8}$ & 8$_{1,7}$    & --3.6663  & 19  & 61.4 &  \ch{CH3C3N} \\ 
114&2866138  & 10$_{1,10}$ & 9$_{1,9}$    & --3.5391  & 21  & 66.8 &  \ch{CH3CH3NH2} \\ 
137&1417100  & 12$_{1,12}$ & 11$_{1,11}$  & --3.3177  & 25  & 79.3 &  \ch{CH3NCO} \\ 
137&7107589  & 12$_{1,11}$ & 11$_{1,10}$  & --3.3143  & 25  & 79.5 &  Unblended \\ 
148&5688153  & 13$_{1,13}$ & 12$_{1,12}$  & --3.3223  & 27  & 86.4 &  aGg$^{\prime}$-\ch{(CH2OH)2}\\ 
149&1852579  & 13$_{1,12}$ & 12$_{1,11}$  & --3.2195  & 27  & 86.6 &  CH$_3$$^{34}$SH and CH$_3$$^{13}$CCH \\ 
\hline 
\noalign{\smallskip}
\end{tabular}
\label{tab:oprot}
\tablefoot{$^{(a)}$ The rotational energy levels are labelled using the conventional notation for asymmetric tops: $J_{K_{a},K_{c}}$, where $J$ denotes the angular momentum quantum number, and the $K_{a}$ and $K_{c}$ labels are projections of $J$ along the $a$ and $c$ principal axes. Lines that were already observed in \citet{Sanz-Novo2024a} are marked with a * symbol.}
\end{table*}

\subsection{Re-inspection of the detection of \ch{HOCS+} towards G+0.693 and search for $K$$_a$ $\geq$ 1 transitions} 
\label{subsec:detection}

The new catalogue, based on the spectroscopy presented in Sect.\,\ref{sec:analysis} and implemented in the Spectral Line Identification and Modeling (SLIM) tool (version from 2023 November 15) within the \textsc{Madcuba} package \citep{martin2019}, was used to conduct a new astronomical search for \ch{HOCS+}. This tool allowed us to generate the local thermodynamic equilibrium (LTE( synthetic spectra under the assumption of LTE conditions.

Once the emission from all the molecules previously identified towards G+0.693 was accounted for (\citealt{Rivilla23} and references therein), we re-evaluated the detection of the $K$$_a$ = 0 lines reported in \citet{Sanz-Novo2024a}, which was based on an extrapolation to higher frequencies of the three lowest-$J$ $R$-branch $a$-type rotational transitions of \ch{HOCS+} measured by \cite{Ohshima96}. \citet{Sanz-Novo2024a} foresaw sizable uncertainties upon reaching the millimetre-wave region (e.g. 0.3 MHz at 100 GHz and 0.6 MHz at 130 GHz, corresponding to velocities of 1.1\,km s$^{-1}$ and 1.8\,km s$^{-1}$, respectively). Nonetheless, we already anticipated that these uncertainties would not have an impact on the analysis, as they are notably smaller than the typical line widths of the molecular line emission observed towards G+0.693 (FWHM $\sim$ 15$-$20 km s$^{-1}$; \citealt{requena-torres_org_2006,requena-torres_lar_2008,zeng2018}). Indeed, the discrepancies between the old and new rest frequencies account for several hundred kilohertz (e.g. 147\,kHz and 360\,kHz at $\sim$103.076\,GHz and $\sim$137.432\,GHz, respectively), which are considerably better than expected and, furthermore, do not affect the detection of any of the previously reported $K$$_a$ = 0 transitions.

Thanks to new laboratory data, we have expanded our search to include $K$$_a$ $\geq$ 1 transitions, which are free from contamination by HNC$^{34}$S (see e.g. the 3$_{1,2}$ -- 2$_{1,1}$ transition at 34.4294313 GHz in Fig.\,\ref{f:LTEspectrum}). In brief, fortuitously, the $B$ and $C$ rotational constants of both HOCS$^+$ and HNC$^{34}$S are extremely similar, and therefore their spectra will be characterised by nearly identical $K_{a}$ = 0 progressions. Therefore, the $K$$_a$ = 0 transitions  of \ch{HOCS+} are partially blended with those belonging to HNC$^{34}$S, although the contribution of the latest can be well constrained based on the HNC$^{32}$S/HNC$^{34}$S isotopic ratio \citep{Sanz-Novo2024a}. Consequently, the inclusion of the newly measured $K_{a} = 1$ transitions is of particular relevance to conclusively disentangle the emission of \ch{HOCS+} and HNC$^{34}$S. Overall, several new $K_a$ = 1 transitions were observed (see Fig. \ref{f:LTEspectrum}), which further strengthen the detection of \ch{HOCS+}, even though they appear to be extremely weak, with most of them contaminated by the emission of other species. Nevertheless, this fact is not surprising owing to the large value of the $A$ rotational constant, which will directly affect the intensity ratio between the $K_{a}$ = 0 and 1 lines.

Regarding the final LTE analysis of \ch{HOCS+}, we used the same parameters of $T_{\rm ex}$ = 28 K, $v$$_{\rm LSR}$ = 66.8 km s$^{-1}$, and FWHM = 21.8 km s$^{-1}$ derived from the previous analysis presented in \citet{Sanz-Novo2024a}. We then carried out the LTE fit to the \ch{HOCS+} emission using the \textsc{Autofit} tool within SLIM \citep{martin2019}, which performs a non-linear least-squares LTE fit to the observed spectra, with only the column density left as a free parameter. 
We used all the $K$$_a$ = 0 transitions covered in our survey, which were already reported in \citet{Sanz-Novo2024a}, as well as the brightest $K$$_a$ = 1 lines listed in Table \ref{tab:oprot}, and accounted for the expected emission from every molecule detected within the same frequency range. Also, since the excitation temperatures of the molecules are low, as usual in G+0.693 ($T_{\rm ex}$\,=\,5$-$20~K), in this case we have not used the vibrational contribution of the partition function ($Q_v$) presented in Table\,\ref{table:partfunct}. We derived a molecular column density of $N$ = (9 $\pm$ 2)\,$\times$\,10$^{12}$\,cm$^{-2}$, which yields a fractional abundance with respect to molecular hydrogen of (7 $\pm$ 2) $\times$ 10$^{-11}$, adopting a $N_{\rm H_2}$ = 1.35\,$\times$\,10$^{23}$\,cm$^{-2}$ from \citet{martin_tra_2008}, and assuming an uncertainty of 15\% of its value. Thus, current results are in perfect agreement with those reported in \citet{Sanz-Novo2024a}. We present the fitted line profiles of the $K_{a}$ = 0 and $K_{a}$ = 1 transitions of \ch{HOCS+} (in red), along with the predicted spectrum considering all molecular species identified and analysed towards G+0.693 (in blue), in Fig. \ref{f:LTEspectrum}(a) and Fig. \ref{f:LTEspectrum}(b), respectively. 

\begin{table}[htbp]
\caption{Rotational and vibrational partition functions for \hocs.}
\label{table:partfunct} \centering
\begin{tabular}{r@{.}lr@{.}lr@{.}l}
\hline\hline   
\noalign{\smallskip}
\multicolumn{2}{c}{\textit{T}[K]}&  \multicolumn{2}{c}{$Q_{rot}$} & \multicolumn{2}{c}{$Q_{vib}^a$} \\
\noalign{\smallskip}
\hline
\noalign{\smallskip}
                 2&725     &    10&26    &   1&00    \\      
                 5&0       &    18&55    &   1&00    \\ 
                 9&375     &    35&73    &   1&00    \\   
                18&75      &    87&32    &   1&00    \\  
                37&5       &   243&07    &   1&00    \\   
                75&0       &   687&32    &   1&00    \\   
\multicolumn{2}{l}{150}    &  1947&00    &   1&02    \\      
\multicolumn{2}{l}{225}    &  3583&92    &   1&11    \\       
\multicolumn{2}{l}{300}    &  5529&52    &   1&27    \\       
\multicolumn{2}{l}{500}    & 11969&00    &   2&08    \\        
\multicolumn{2}{l}{1000}   & 34026&56    &   8&01    \\         
\hline
\end{tabular}
\tablefoot{
$^a$Anharmonic vibrational energies taken from \citet{Fortenberry2012}.}
\end{table}

\section{Conclusions}

In this study, we conducted laboratory experiments aimed at extending the high-resolution spectroscopy of protonated carbonyl sulphide (\hocs), providing a comprehensive database for its full astronomical characterisation well into the submillimetre wave region. Here, we summarise the key findings and implications of our research:

\begin{itemize}
    \item Utilising CASAC, we employed a frequency-modulated free-space absorption spectrometer for our experiments. Despite extensive efforts and optimisations, weak signals persisted, necessitating prolonged averaging to achieve acceptable signal-to-noise ratios. Temperature stability emerged as a critical factor influencing plasma behaviour, with even slight fluctuations affecting discharge efficacy and magnetic field performance.
    \item Our experiments have successfully detected rotational transitions of \hocs\, in the laboratory setting. Through careful analysis and baseline subtraction techniques, we obtained a dataset comprising 30 rotational transitions, with a standard deviation of 46 kHz. The accuracy of our dataset is estimated to be within the range of 75$-$150\,kHz ($\sim$\,100\,m/s at 300\,GHz).
    \item Comparisons of our experimental results with prior infrared measurements and quantum chemical calculations reveal an overall agreement in spectroscopic parameters. While discrepancies are noted in certain quartic parameters, most rotational constants exhibit good consistency across studies.
    \item The new spectroscopic measurements reported in this work allow us to decisively corroborate the detection of the  $K_a$ = 0 lines presented in \citet{Sanz-Novo2024a}, thanks to more accurate constraints on both the $B$ and $C$ rotational constants as well as the centrifugal distortion parameters. Also, we have found additional $K_a$ = 1 lines to unravel the emission of \hocs\, and HNC$^{34}$S, which is consistent with the detection of the $K_a$ = 0 lines alone, strengthening the detection even further.
    \item Our laboratory measurements also provide crucial insights for future astronomical observations of \hocs\, in other interstellar environments, such as the cold sources TMC-1 or L1544, where the line widths can be as narrow as 0.5$-$0.6 km s$^{-1}$ (see e.g. the QUIJOTE line survey; \citealt{Cernicharo21sulfur}).    
    \item Future laboratory research endeavours may focus on refining experimental techniques to enhance sensitivity and resolution, and possibly to pin down the $b$-type spectrum of the cation, still undetected, which would further constrain the $A$ rotational constant. 
\end{itemize}

\begin{acknowledgements}

We gratefully acknowledge the Max Planck society for the financial support. V.M.R., M.S-N. I.J-S. and J.M.P. acknowledge support from the grant No. PID2022-136814NB-I00 by the Spanish Ministry of Science, Innovation and Universities/State Agency of Research MICIU/AEI/10.13039/501100011033 and by ERDF, UE. V.M.R. also acknowledges support from the grant number RYC2020-029387-I funded by MICIU/AEI/10.13039/501100011033 and by "ESF, Investing in your future", and from the Consejo Superior de Investigaciones Cient{\'i}ficas (CSIC) and the Centro de Astrobiolog{\'i}a (CAB) through the project 20225AT015 (Proyectos intramurales especiales del CSIC), and from the grant CNS2023-144464 funded by MICIU/AEI/10.13039/501100011033 and by \lq\lq European Union NextGenerationEU/PRTR''. M.S.N. also acknowledges a Juan de la Cierva Postdoctoral Fellowship, project JDC2022-048934-I, funded by MCIN/AEI/10.13039/501100011033 and by the European Union “NextGenerationEU/PRTR”

\end{acknowledgements}

\bibliographystyle{aa}
\bibliography{HOCS+,Miguel_b}

\begin{appendix}
\section{Observed \ch{HOCS+} transitions towards G+0.693–0.027}

In this section the spectra of the $K$$_a$ = 0 and $K$$_a$ = 1 transitions detected towards the molecular cloud G+0.693–0.027 are reported.

\begin{center}
\begin{figure*}[ht]
     \centerline{\resizebox{0.82\hsize}{!}{\includegraphics[angle=0]{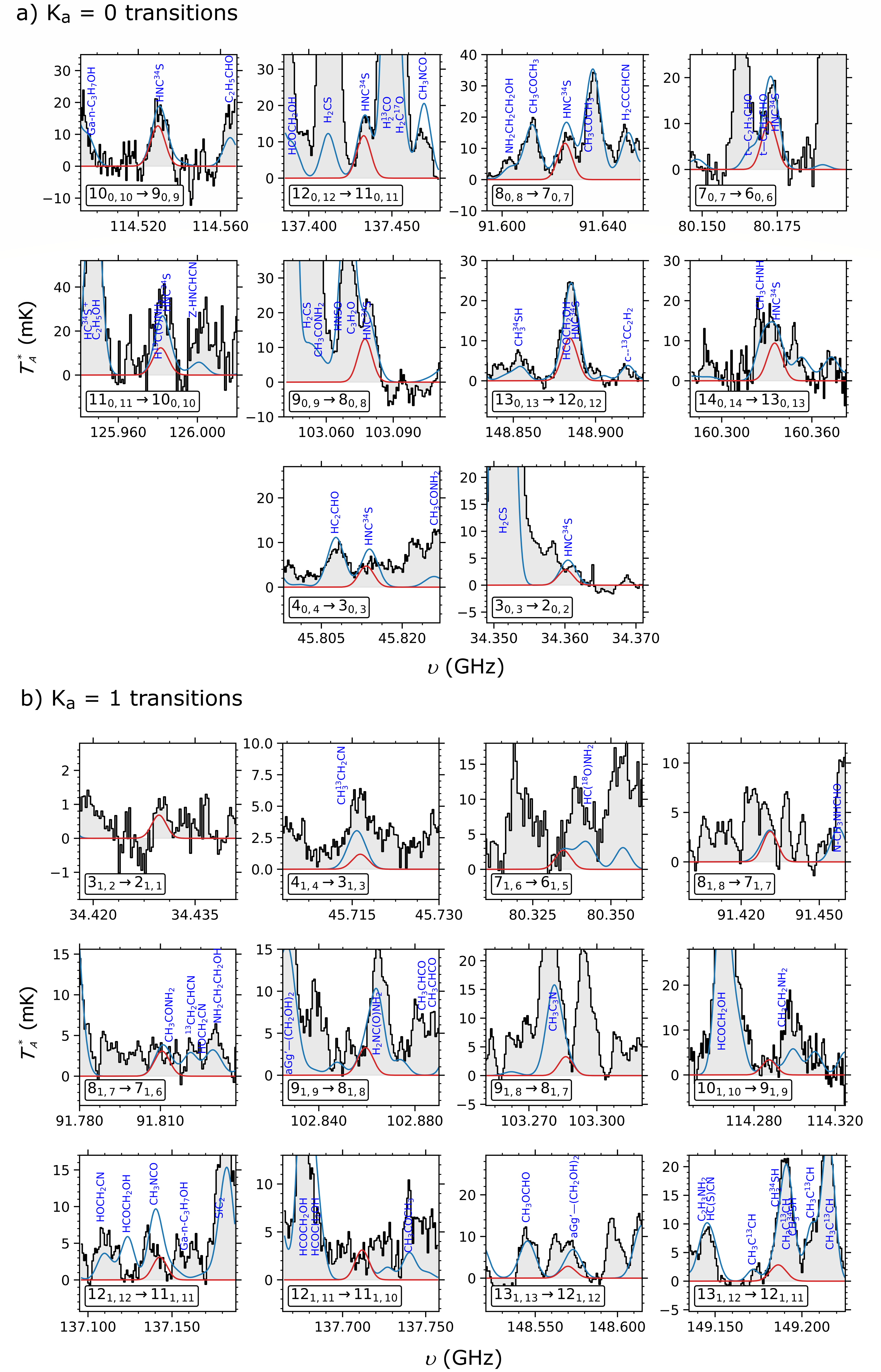}}}
     \caption{Transitions of \ch{HOCS+} detected towards the G+0.693–0.027 molecular cloud (listed in Table \ref{tab:oprot}). We depict with a red line the result of the best LTE fit of \ch{HOCS+}, while the blue line plots the emission from all the molecules identified to date in our survey. The observed spectra are plotted as gray histograms. Note that for the $K$$_a$ = 0 transitions, \ch{HOCS+} appears to be blended with HNC$^{34}$S because of their similar $B$ + $C$, but the contribution of HNC$^{34}$S can be well constrained based on the HNC$^{32}$S/HNC$^{34}$S isotopic ratio \citep{Sanz-Novo2024a}.}
\label{f:LTEspectrum}
\end{figure*}
\end{center}

\end{appendix}

\end{document}